\documentclass[prb,twocolumn,superscriptaddress,showpacs]{revtex4}
\usepackage{amsmath}
\usepackage{graphicx}
\usepackage{dcolumn}
\usepackage{bm}
\usepackage{multirow}

\begin{document}

\title{Evaluation of Born and local effective charges in unoriented materials from vibrational spectra}
\author{X.S. Xu}
\affiliation{Department of Chemistry, University of Tennessee, Knoxville, TN 37996}
\author{Q.-C. Sun}
\affiliation{Department of Chemistry, University of Tennessee, Knoxville, TN 37996}
\author{R. Rosentsveig}
\affiliation{Department of Materials and Interfaces, Weizmann Institute of Science, RehoVot 76100, Israel}
\author{J.L. Musfeldt}
\affiliation{Department of Chemistry, University of Tennessee, Knoxville, TN 37996}

\date{\today }

\begin{abstract}
We present an application of the Lorentz model in which fits to vibrational
spectra or a Kramers Kronig analysis are employed along with several useful
formalisms to quantify microscopic charge in unoriented (powdered)
materials. 
The conditions under which these techniques can be employed are discussed,
and we analyze the vibrational response of a layered transition metal
dichalcogenide and its nanoscale analog to illustrate the utility 
of this approach. 

%
\end{abstract}


\maketitle

\narrowtext

\clearpage

\section{Introduction}

Born and local (or ionic) effective charge are well-known quantities with
which to assess chemical bonding and polarization in a material.\cite{Born1918,Szigeti1950} A
large local effective charge indicates a highly ionic system (for instance,
0.8 in NaCl), 
a medium-sized value points toward an intermediate bonding case (0.4 in
GaAs), and a small value is associated with a covalent material (0 in Si).%
\cite{Ashcroft1976, Kittel1966} 
On the other hand, Born effective charge describes the static and dynamic polarizations.\cite{Born1954}
From the optical
properties point of view, large longitudinal optic-transverse optic (LO-TO)
splittings are well-known characteristics of polarizable compounds and
associated with substantial Born effective charges. This is because LO-TO
splitting is directly proportional to charge within Born's original
formalism.\cite{Born1918} Using the LO-TO splitting and high frequency
dielectric constant, it is relatively straightforward to quantify chemical
bonding 
from optical measurements of high quality single crystal samples and compare
the extracted value(s) with first principles calculations.\cite%
{Lee2003,Fontana1984,Zhong1994} Unfortunately, there are many instances when
single crystals of a bulk material are unavailable, either because they can
not be grown or are not of sufficient size or quality for optical
measurements. At the same time, nanoparticles, nanotubes, and alloys (or
composite mixtures) present scientifically compelling problems, \cite%
{Long2007,Tenne2004,Poudel2008,Hochbaum2008,Boukai2008,Cao2007,Carr1985}
where optical measurements on powdered materials are the only
option, a drawback that complicates the situation but does not
diminish the desirability of obtaining quantitative Born and local
effective charge data.

Following Born and Szigeti,\cite{Born1918,Szigeti1950} we present an
application of the Lorentz model in which fits to vibrational spectra or a
Kramers Kronig analysis are employed along with several useful formalisms to
quantify microscopic charge in unoriented (powdered) materials, assuming
that the effects of ionic displacement and atomic polarizability are
superimposable. 
We demonstrate that this technique can be used to assess chemical bonding
and local strain under certain conditions, a development that advances the
field of nanoscience and, at the same time, retains many attractive features
of optical spectroscopy and the traditional Lorentz model. This paper is
organized as follows. Sections II and III review the situation for single
and multiple collinear oscillators. Sections IV and V present how this
approach must be modified for a randomly oriented (powder) sample. Section
VI illustrates the use of this technique to quantify Born and local (or
ionic) charge in powdered 2H-MoS$_{2}$ and its nanoscale analog. Comparison
of our analysis of the powdered 2H-MoS$_{2}$ data with that of the single
crystal shows that this approach accounts almost perfectly for sample
orientation. \cite{Sun2009,Wieting1971,Uchida1978} The extension to assess
size effects in nanoparticle samples demonstrates its utility.\cite{Sun2009}
Our objective is to clearly present the framework, useful equations, and an
application of this method. The Appendix provides alternate frameworks of
the model.

%
%

\section{Collinear oscillator}

\subsection{A Review of the General Formalism}

\label{MarkerGeneral}

Starting with the Lorentz model for bound charge,\cite{Wooten1972,Born1954}
one has:

\begin{equation}
m\overset{\cdot \cdot }{x}+m\gamma \overset{\cdot }{x}+m\omega _{0}^{2}x=qE,
\label{EqLorentzmo}
\end{equation}%
where $x$ is the relative displacement of positive and negative ions, $q$ is
the effective ionic charge, $m$ is the reduced mass, $E$ is the electric
field, $\gamma $ is the damping parameter and $\omega _{0}$ is the
oscillator frequency. This model is applicable to vibrational modes of
isolated oscillators such as those in gas phase molecules. Due to
dipole-dipole interactions between oscillators in a solid, so called
depolarization effects must be considered. Here, $E$ has to be replaced by
the microscopic field $E_{mic}:$\cite{Huang1988}

\begin{equation}
E\rightarrow E_{mic}=E+\eta \frac{P}{\epsilon _{0}} ,  \label{EFields}
\end{equation}%
where $E_{mic}$ is the microscopic field that is really acting on the
oscillator, $E$ is the macroscopic field that should be used to calculate
the dielectric constant, $P$ is the polarization, and $\eta $ is the
depolarization factor which depends on the topological arrangement of the
oscillators. With this substitution, Eq. (\ref{EqLorentzmo}) becomes:

\begin{equation}
m\overset{\cdot \cdot }{x}+m\gamma \overset{\cdot }{x}+m\omega
_{0}^{2}x=q(E+\eta \frac{P}{\epsilon _{0}}).  \label{EqLorentzso}
\end{equation}

The overall goal is to find the dielectric constant which relates $E$ and $P$%
. However, in Eq. (\ref{EqLorentzso}), there are 3 unknown variables: $x$, $%
E $ and $P$. One more equation is needed to find the relationship between $E$
and $P$. This additional information will come from the definition of $P$.
Assuming the polarization $P$ is a linear combination of ionic contributions
$P_{i}=\frac{xq}{V}$, which comes from relative displacements of ions, and
electronic contributions $P_{e}=\frac{\epsilon _{0}\alpha }{V}E_{mic}$ which
is due to the distortion of electron clouds, one has

\begin{eqnarray*}
P &=&P_{i}+P_{e} \\
&=&\frac{xq}{V}+\frac{\epsilon _{0}\alpha }{V}E_{mic},
\end{eqnarray*}%
where $\alpha $ and $V$ are the polarizability and volume of the oscillator,
respectively. Substituting the expression for $E_{mic}$ in Eq. (\ref{EFields}%
), one has the second important equation:

\begin{equation*}
P=\frac{xq}{V}+\frac{\epsilon _{0}\alpha }{V}E+\frac{\eta \alpha }{V}P,
\end{equation*}%
or

\begin{equation}
P=\frac{1}{1-\eta \alpha /V}(\frac{xq}{V}+\frac{\epsilon _{0}\alpha }{V}E).
\label{EqPolar}
\end{equation}

\noindent Taken together, Eqs. (\ref{EqLorentzso}) and (\ref{EqPolar}) are
enough for dealing with single crystal problems (where there is
translational symmetry). 
Next we discuss the dielectric response.

To obtain the expression for the dielectric constant which relates $P$ and $%
E $, we must eliminate $P$ From Eq. (\ref{EqLorentzso}). Plugging Eq. (\ref%
{EqPolar}) into Eq. (\ref{EqLorentzso}), one has%

\begin{equation*}
m\overset{..}{x}+m\gamma \overset{\cdot }{x}+m\omega _{0}^{2}x=q[E+\frac{%
\eta }{\epsilon _{0}}\frac{1}{1-\eta \alpha /V}(\frac{xq}{V}+\frac{\epsilon
_{0}\alpha }{V}E)],
\end{equation*}%
or

\begin{equation*}
m\overset{\cdot \cdot }{x}+m\gamma \overset{\cdot }{x}+[m\omega _{0}^{2}-%
\frac{q^{2}\eta }{\epsilon _{0}V(1-\eta \alpha /V)}]x=\frac{qE}{1-\eta
\alpha /V}.
\end{equation*}


\noindent For $x=x_{0}e^{-i\omega t}$, the solution is

\begin{equation}
x=\frac{\frac{q}{m(1-\eta \alpha /V)}E}{[\omega _{0}^{2}-\frac{q^{2}\eta }{%
m\epsilon _{0}V(1-\eta \alpha /V)}]-\omega ^{2}-i\gamma \omega }.
\label{Eqx1}
\end{equation}

\noindent For simplicity, we define:%
\begin{equation*}
\omega _{1}^{2}\equiv \frac{q^{2}\eta }{m\epsilon _{0}V(1-\eta \alpha /V)}.
\end{equation*}

Using Eq. (\ref{EqPolar}), we can write down the polarization:%
\begin{eqnarray}
P &=&\frac{1}{1-\eta \alpha /V}(\frac{xq}{V}+\frac{\epsilon _{0}\alpha }{V}E)
\notag \\
&=&\frac{1}{1-\eta \alpha /V}[\frac{q}{V}\frac{\frac{q}{m(1-\eta \alpha /V)}E%
}{\omega _{0}^{2}-\omega _{1}^{2}-\omega ^{2}-i\gamma \omega }+\frac{%
\epsilon _{0}\alpha }{V}E]  \notag \\
&=&\frac{\epsilon _{0}\alpha /V}{1-\eta \alpha /V}E  \notag \\
&&+\frac{q^{2}}{mV(1-\eta \alpha /V)^{2}}\frac{E}{\omega _{0}^{2}-\omega
_{1}^{2}-\omega ^{2}-i\gamma \omega }.  \label{EqPolar1}
\end{eqnarray}

Therefore, the dielectric constant is
\begin{eqnarray}
\varepsilon  &=&1+\frac{P}{\epsilon _{0}E}  \notag \\
&=&1+\frac{\alpha /V}{1-\eta \alpha /V}  \notag \\
&&+\frac{q^{2}}{\epsilon _{0}mV(1-\eta \alpha /V)^{2}}\frac{1}{\omega
_{0}^{2}-\omega _{1}^{2}-\omega ^{2}-i\gamma \omega }.  \label{Eqdi}
\end{eqnarray}%
We can immediately see that the Lorentz model in a solid is modified
compared with that of an isolated oscillator due to the depolarization
effect and the polarizability (when $\alpha =0$ and $\eta =0$, Eq. (\ref%
{Eqdi}) reduces to the case of an isolated
oscillator.\cite{Wooten1972}). We end this section by summarizing
several useful definitions and expressions that connect measurable
quantities (left-hand side) to microscopic parameters (right-hand
side). These include the high-frequency dielectric constant,
oscillator strength, and TO phonon frequency.
\begin{subequations}
\begin{eqnarray}
\varepsilon (\infty ) &\equiv &1+\frac{\alpha /V}{1-\eta \alpha /V}
\label{EqExpTh} \\
A &\equiv &\frac{q^{2}}{\epsilon _{0}mV(1-\eta \alpha /V)^{2}}
\label{EqExpTh1} \\
\omega _{TO}^{2} &\equiv &\omega _{0}^{2}-\omega _{1}^{2}  \notag \\
&=&\omega _{0}^{2}-\frac{q^{2}\eta }{\epsilon _{0}mV(1-\eta \alpha /V)}
\end{eqnarray}%
Clearly, one can extract $\varepsilon (\infty )$, $A$, and $\omega
_{TO}^{2}$ from the experimental vibrational spectra using an
oscillator fit or Kramers Kronig analysis and use these quantities
to extract the microscopic parameters.\cite{Wooten1972} In other
words, once $\varepsilon (\infty )$ and $A$ are
measured, one can derive $\alpha $ and $q$ using Eqs. (\ref{EqExpTh}) and (%
\ref{EqExpTh1}). Note that
\end{subequations}
\begin{equation*}
S\equiv \frac{A}{\omega _{TO}^{2}}
\end{equation*}%
is the dimensionless oscillator strength.

\subsection{Evaluating the Effective Charges}

\label{MarkerEffcharge}


Ionic effective charge (also called local effective charge) $q$ is an
important quantity because it quantifies the ionicity. It is distinct from
the Born effective charge as discussed below. The most straightforward way
to extract ionic effective charge from oscillator strength is to use Eq. (%
\ref{EqExpTh1}):

\begin{equation}
q^{2}=\epsilon _{0}mV(1-\eta \alpha /V)^{2}A.
\end{equation}

\noindent Thus, quantitative information about bond covalency/ionicity can
be extracted from a knowledge of oscillator strength and polarizability.


In the literature, we often find Born effective charge defined as:\cite%
{Gonze1997}

\begin{equation}
\frac{q_{B}^{2}}{\epsilon _{0}mV}=\varepsilon (\infty )(\omega
_{LO}^{2}-\omega _{TO}^{2}),
\end{equation}%
which, as shown in the Appendix, is equivalent to:%
\begin{equation}
\frac{q_{B}^{2}}{\epsilon _{0}mV}=A=\frac{q^{2}}{\varepsilon _{0}mV(1-\eta
\alpha /V)^{2}}.
\end{equation}%
%
%
Therefore Born and local ionic charge are related as
\begin{equation}
q_{B}^{2}=\frac{q^{2}}{(1-\eta \alpha /V)^{2}}.  \label{EqChargerel}
\end{equation}%
Note that Born effective charge takes into account the combined
contributions of ionic displacement, electron cloud deformation, and
depolarization effects, whereas ionic effective charge only represents the
charge of the ions.%
%


\section{Multiple Collinear Oscillators}

When there is more than one oscillator (as in most real materials),
one has to go back to the polarizability (Eq. (\ref{EqPolar1})) and
add a mode index $j$. This yields:

\begin{eqnarray}
P_{j} &=&\frac{\alpha _{j}/V}{1-\eta _{j}\alpha _{j}/V}E  \notag
\label{EqPolar2} \\
&&+\frac{q_{j}^{2}}{m_{j}V(1-\eta _{j}\alpha _{j}/V)^{2}}\frac{E}{\omega
_{0}^{2}-\omega _{1,j}^{2}-\omega ^{2}-i\gamma _{j}\omega }  \notag \\
\end{eqnarray}%
and
\begin{equation*}
\omega _{1,j}^{2}\equiv \frac{q_{j}^{2}\eta _{j}}{\epsilon _{0}m_{j}V(1-\eta
_{j}\alpha _{j}/V)}.
\end{equation*}

Then the dielectric constant and other definitions can be expanded as:%
\begin{eqnarray}
\varepsilon  &=&1+\frac{\underset{j}{\sum }P_{j}}{\epsilon _{0}E}  \notag \\
&=&1+\underset{j}{\sum }\frac{\alpha _{j}/V}{1-\eta _{j}\alpha _{j}/V}
\notag \\
&&+\underset{j}{\sum }\frac{q_{j}^{2}}{\epsilon _{0}m_{j}V(1-\eta _{j}\alpha
_{j}/V)^{2}}\frac{1}{(\omega _{0}^{2}-\omega _{1,j}^{2})-\omega ^{2}-i\gamma
_{j}\omega }  \notag \\
&=&1+\underset{j}{\sum }[\varepsilon _{j}(\infty )-1]+\underset{j}{\sum }%
A_{j}\frac{1}{\omega _{TO,j}^{2}-\omega ^{2}-i\gamma _{j}\omega },
\end{eqnarray}%
where
\begin{subequations}
\begin{eqnarray}
\varepsilon _{j}(\infty ) &=&1+\frac{\alpha _{j}/V}{1-\eta _{j}\alpha _{j}/V}
\\
A_{j} &=&\frac{q_{j}^{2}}{\epsilon _{0}m_{j}V(1-\eta _{j}\alpha _{j}/V)^{2}}
\\
\omega _{TO,j}^{2} &=&\omega _{0,j}^{2}-\omega _{1,j}^{2}  \notag \\
&=&\omega _{0,j}^{2}-\frac{q_{j}^{2}\eta _{j}}{\epsilon _{0}m_{j}V(1-\eta
_{j}\alpha _{j}/V)}.
\end{eqnarray}

\noindent Note that polarizability represents the high frequency dielectric
response of the electron cloud. Therefore, $\alpha $ should be labeled
according to the polarization direction, although for simplicity, we label
this quantity with the mode index.

\section{Tilted oscillator}

If the electric field is not perfectly aligned with the direction of motion
of a certain mode, the observed oscillator strength will be reduced from its
intrinsic value. This can be easily understood by considering the extreme
case: when the light is polarized perpendicular to a particular mode, this
mode will not contribute to the oscillator strength at all. Thus, if one
directly employs the formulas for local ionic and Born effective charge in
Section \ref{MarkerEffcharge} without modification, the results will be
unphysical. This is because a tilted oscillator provides only a component of
the total intrinsic oscillator strength. Instead, we must go back to Section %
\ref{MarkerGeneral} and rederive a set of formulas that take orientation
into account.

We can employ a modified version of Eq. (\ref{Eqx1}) to account for the
effect of a tilted oscillator:

\end{subequations}
\begin{equation}
x(\theta )=\frac{\frac{q}{m(1-\eta \alpha /V)}E\cos (\theta )}{\omega
_{0}^{2}-\omega _{1}^{2}-\omega ^{2}-i\gamma \omega }.
\end{equation}%
Hence,
\begin{eqnarray}
P(\theta ) &=&\frac{1}{1-\eta \alpha /V}[\frac{xq}{V}+\frac{\epsilon
_{0}\alpha }{V}E\cos (\theta )]  \notag \\
&=&\frac{\epsilon _{0}\alpha /V}{1-\eta \alpha /V}E\cos (\theta )  \notag \\
&&+\frac{q^{2}}{mV(1-\eta \alpha /V)^{2}}\frac{E\cos (\theta )}{\omega
_{0}^{2}-\omega _{1}^{2}-\omega ^{2}-i\gamma \omega },  \label{EqPolar3}
\end{eqnarray}%
and%
\begin{eqnarray}
\varepsilon (\theta ) &=&1+\frac{P(\theta )\cos (\theta )}{\epsilon _{0}E}
\notag \\
&=&1+\frac{\alpha /V\cos ^{2}(\theta )}{1-\eta \alpha /V}  \notag \\
&&+\frac{q^{2}}{\epsilon _{0}mV(1-\eta \alpha /V)^{2}}\frac{\cos ^{2}(\theta
)}{\omega _{0}^{2}-\omega _{1}^{2}-\omega ^{2}-i\gamma \omega }.  \notag \\
\end{eqnarray}%
Here, $\theta $ is the angle between the electric field and the direction in
which oscillator intensity is maximum. If a measurement is done on a sample
with a distribution of orientations (\{$\theta$\}), the result is
appropriately averaged as:
\begin{eqnarray}
\widetilde{\varepsilon } &=&\langle \varepsilon (\theta )\rangle  \notag \\
&=&1+\frac{\alpha /V\langle \cos ^{2}(\theta )\rangle }{1-\eta \alpha /V}
\notag \\
&&+\frac{q^{2}}{\epsilon _{0}mV(1-\eta \alpha /V)^{2}}\frac{\langle \cos
^{2}(\theta )\rangle }{\omega _{0}^{2}-\omega _{1}^{2}-\omega ^{2}-i\gamma
\omega }.
\end{eqnarray}%
In this case, the observed parameters are related to the microscopic
parameters as:
\begin{subequations}
\begin{eqnarray}
\widetilde{\varepsilon }(\infty ) &=&1+\frac{\alpha /V\langle \cos
^{2}(\theta )\rangle }{1-\eta \alpha /V} \\
\widetilde{A} &=&\frac{q^{2}}{\epsilon _{0}mV(1-\eta \alpha /V)^{2}}\langle
\cos ^{2}(\theta )\rangle \\
\widetilde{\omega }_{TO}^{2} &=&\omega _{0}^{2}-\omega _{1}^{2}  \notag \\
&=&\omega _{0}^{2}-\frac{q^{2}\eta }{\epsilon _{0}mV(1-\eta \alpha /V)}.
\end{eqnarray}
Here, the brackets $\langle\rangle$ indicate directional averaging.

\section{Multiple tilted oscillators}

Most isotropic samples of real materials have several vibrational modes. We
can write down the dielectric constant for the case of multiple tilted
oscillators by combining Eqs. (\ref{EqPolar2}) and (\ref{EqPolar3}):

\end{subequations}
\begin{align}
\varepsilon (\{\theta _{j}\})& =1+\underset{j}{\sum }\frac{P(\theta
_{j})\cos (\theta _{j})}{\epsilon _{0}E} \\
& =1+\underset{j}{\sum }\frac{\alpha _{j}/V}{1-\eta _{j}\alpha _{j}/V}\cos
^{2}(\theta _{j})  \notag \\
& +\underset{j}{\sum }\frac{q_{j}^{2}\cos ^{2}(\theta _{j})}{\epsilon
_{0}m_{j}V(1-\eta _{j}\alpha _{j}/V)^{2}}\frac{1}{(\omega _{0}^{2}-\omega
_{1,j}^{2})-\omega ^{2}-i\gamma _{j}\omega },
\end{align}%
and%
\begin{equation*}
\varepsilon (\{\theta _{j}\})=1+\underset{j}{\sum }[\varepsilon _{j}(\infty
)-1]\cos ^{2}(\theta _{j})+\underset{j}{\sum }A_{j}\frac{\cos ^{2}(\theta
_{j})}{\omega _{TO}^{2}-\omega ^{2}-i\gamma \omega }.
\end{equation*}

\noindent Note that the \{$\theta _{j}$\} are related to an
oscillator's polarization direction. Therefore, the number of
independent $\theta _{j}$ may be less than the number of modes.
Hence,

\begin{eqnarray}
\widetilde{\varepsilon } &=&\langle \varepsilon (\{\theta _{j}\})\rangle
\notag \\
&=&1+\underset{j}{\sum }[\varepsilon _{j}(\infty )-1]\langle \cos
^{2}(\theta _{j})\rangle   \notag \\
&+&\underset{j}{\sum }A_{j}\frac{\langle \cos ^{2}(\theta _{j})\rangle }{%
\omega _{TO,j}^{2}-\omega ^{2}-i\gamma _{j}\omega }  \notag \\
&=&\widetilde{\varepsilon }(\infty )+\underset{j}{\sum }\widetilde{A_{j}}%
\frac{1}{\omega _{TO,j}^{2}-\omega ^{2}-i\gamma _{j}\omega },
\label{Eqepsf1}
\end{eqnarray}%
where the observed (apparent) parameters are:

\begin{subequations}
\begin{eqnarray}
\widetilde{\varepsilon }(\infty ) &=&1+\underset{j}{\sum }\frac{\alpha
_{j}/V\langle \cos ^{2}(\theta _{j})\rangle }{1-\eta _{j}\alpha _{j}/V}
\notag \\
&=&1+\underset{j}{\sum }[\varepsilon _{j}(\infty )-1]\langle \cos
^{2}(\theta _{j})\rangle   \label{eps1} \\
\widetilde{A_{j}} &=&\frac{q_{j}^{2}}{\epsilon _{0}m_{j}V(1-\eta _{j}\alpha
_{j}/V)^{2}}\langle \cos ^{2}(\theta _{j})\rangle   \label{eps2} \\
\widetilde{\omega }_{TO,j}^{2} &=&\omega _{0,j}^{2}-\omega _{1,j}^{2}  \notag
\\
&=&\omega _{0,j}^{2}-\frac{q_{j}^{2}\eta _{j}}{\epsilon _{0}m_{j}V(1-\eta
_{j}\alpha _{j}/V)}.  \label{eps3}
\end{eqnarray}

If all vibrational features can be resolved in frequency space, $\widetilde{%
\varepsilon}(\infty )$ and $\widetilde{A_{j}}$ can be determined from a
model oscillator fit or a Kramers-Kronig analysis. At the same time, the $%
\eta _{j}$ can be estimated from the crystal structure.\cite{Huang1988}
However, one can not use the $N+1$ equations given by Eqs. (\ref{eps1}, \ref%
{eps2}) to solve for $3N$ unknowns. The latter include $\alpha _{j}$, $q_{j}$%
, and $\langle \cos ^{2}(\theta _{j})\rangle $. Even if in some
cases, we know $\langle \cos ^{2}(\theta _{j})\rangle $ (perhaps
from an independent x-ray measurement), there are still $2N$
unknowns to determine from only $N+1 $ equations. Additional
information is needed to constrain the system.

Despite this limitation, vibrational spectroscopy of unoriented
powdered samples can still be an important tool for extracting
microscopic charge and bonding information. There are two important
cases:

\textbf{Case 1:} The system is simple enough to have $N+1=2N$. For
example, in rocksalts, the crystal symmetry is cubic ($\langle\cos
^{2}(\theta
_{j})\rangle$ is always 1/3), and there is only one vibrational mode ($N=1$%
). Thus powder spectroscopy is sufficient to determine all of the
microscopic parameters for systems such as NaCl or MnO.

\textbf{Case 2:} Occasionally, some variables are already known,
say, from another method, sample, or similar compound, so that the
total number of known variables can be reduced to be equal to or
less than $N+1$. Recent work on MoS$_{2}$ nanoparticles provides a
good example.\cite{Sun2009} Here, the interplane oscillator strength
is identical to that of the single crystal. Therefore, it is
reasonable to assume that the corresponding polarizability and
charge are both the same for the nanoparticles as they are in the
single crystal, a coincidence that  reduces the number of unknowns
and makes the extraction of intra-plane charge bonding information
possible. We elaborate on the case of MoS$_2$ below.

\section{The dynamics of a model transition metal dichalcogenide: testing
our approach}

In order to test the workability of this approach, we elected to investigate
a model transition metal dichalcogenide. The bulk material, 2H-MoS$_{2}$,
belongs to the $P6_{3}/mmc$ space group (Fig. \ref{fig1}(a)).\cite%
{Schonfeld1983} One consequence of this layered architecture is the
low-dimensional electronic structure which consists of strong bonding within
layers 
and weak van der Waals interactions between layers.\cite{Verble1970} Each MoS%
$_{2}$ slab contains a layer of metal centers, sandwiched between two
chalcogen layers, with each metal atom bonded to six chalcogen atoms in a
trigonal prismatic arrangement. There are two infrared active $E_{1u}$ and $%
A_{2u}$ vibrational modes.\cite{Verble1970} Schematic views of these
displacement patterns are shown in Fig. \ref{fig1}(a). The $E_{1u}$ and $%
A_{2u}$ symmetries characterize intralayer and interlayer motions,
respectively. 
We begin with demonstrating the self-consistency of the theory by analyzing
the oscillator orientation and predicting the observed optical parameters.
We then extend our technique to the chemically-identical but morphologically
different nanoparticles to illustrate the consequences of finite size,
strain, and curvature.

\begin{figure}[h]
\centerline{\includegraphics[width = 2.32in]{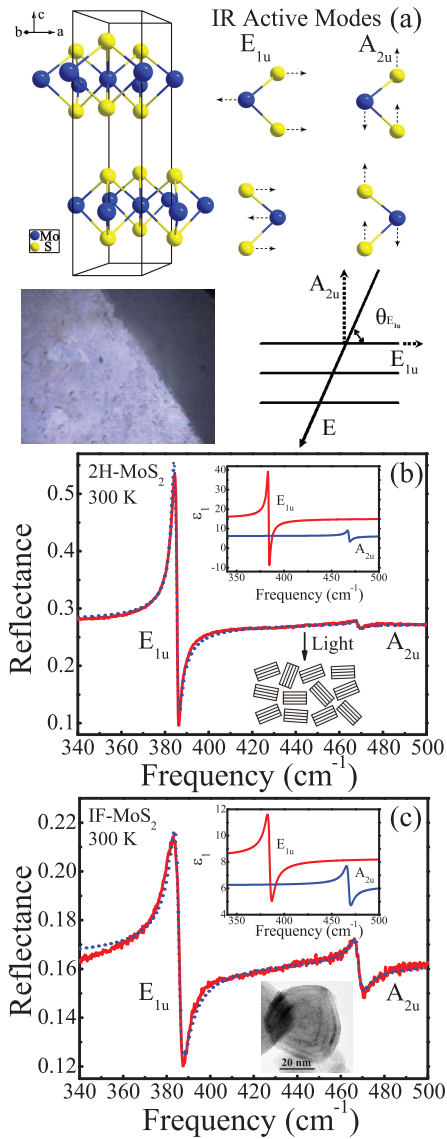}}
\caption{(Color online) (a) Crystal structure of
2H-MoS$_2$,\protect\cite{Schonfeld1983} schematic view of the
displacement patterns of the infrared-active modes, photo of a
typical pressed pellet sample, and diagram for tilted oscillators.
(b) Close-up view of the 300 K reflectance spectra of 2H-MoS$_2$
powder. The red solid curve is experimental data and the blue dashed
line is a fit according to Eqs. (\protect\ref{EqR}) and
(\protect\ref{Eqdi2}). The insets show  the dielectric constant and
a schematic view of the 2H-MoS$_2$ platelets in a pressed pellet
sample. (c) Close-up view of the 300 K reflectance of IF-MoS$_2$. An
identical color scheme and line type is employed. The insets show
the dielectric response and a high resolution TEM image of an
IF-MoS$_2$ nanoparticle. } \label{fig1}
\end{figure}

\begin{table}[tbh]
\caption{Apparent parameters extracted from an oscillator fit to the
measured reflectance spectrum of 2H-MoS$_{2}$ powder, and the intrinsic
parameters from a single crystal sample. \protect\cite{Wieting1971}}%
\begin{ruledtabular}
\begin{tabular}{l|ll|cl}
&$\widetilde{S}_j$ \footnotemark[1]& $ \widetilde{\epsilon}(\infty)$  \footnotemark[1] &$S_j$\footnotemark[2]& $\epsilon(\infty)$\footnotemark[2]\\

\hline
$E_{1u}$ &0.114  &\multirow {2}{*}{10.3} &0.20  & 15.2   \\
$A_{2u}$&0.0036 &&0.030 &  6.2 \\
\end{tabular}
\footnotetext[1]{Apparent parameters obtained from an oscillator
fitting analysis of the measured powder spectrum.}
\footnotetext[2]{Intrinsic parameters obtained from a fit to the
measured reflectance of a single crystal sample.\cite{Wieting1971}}
\label{Tab1}
\end{ruledtabular}
\end{table}

Figure \ref{fig1}(b) displays a close-up view of the reflectance of 2H-MoS$%
_{2}$.\cite{Experiments} As expected for a pressed powder sample, both the $%
E_{1u}$ and $A_{2u}$ modes are clearly observed. At normal incidence, the
dielectric function is related to reflectance as:
\end{subequations}
\begin{equation}
R(\omega) = {\displaystyle{\left| \frac{\sqrt{\epsilon(\omega)}-1}{\sqrt{%
\epsilon(\omega)}+1} \right| }}^2.  \label{EqR}
\end{equation}

\noindent Equation (\ref{EqR}) is formally valid for single-bounce
reflectance at the interface of 2 semi-infinite media. For the case
of real materials with finite thickness, sample thickness must be
sufficient to assume that there is no back reflectance. Large
attenuation due to a strong resonance is helpful here.

A fit to the 2H-MoS$_{2}$ powder data using Eqs. (\ref{Eqepsf1}) and (\ref%
{EqR}) (Fig. \ref{fig1}(b)) allows us to extract $\widetilde{\varepsilon }%
(\infty )=10.3$, $\widetilde{S}_{E_{1u}}=0.114$ and $\widetilde{S}%
_{A_{2u}}=0.0036$ (Table I). We refer to these values as ``apparent
oscillator parameters" because they are obtained from direct fits to
the measured powder spectrum. They are distinct from the
\textquotedblleft
intrinsic parameters" 
that are useful for evaluation of Born and ionic charge.

The apparent and intrinsic oscillator parameters are related according to
Eqs. (\ref{eps1}- \ref{eps3}) by the oscillator orientation. For MoS$_{2}$,
one has:

\begin{subequations}
\begin{eqnarray}
\widetilde{\varepsilon }(\infty ) &=&1+[\varepsilon _{E_{1u}}(\infty
)-1]\langle \cos ^{2}(\theta _{E_{1u}})\rangle v  \notag \\
&+&[\varepsilon _{A_{2u}}(\infty )-1]\langle \cos ^{2}(\theta
_{A_{2u}})\rangle v  \label{Eqsmp1} \\
\widetilde{S}_{E_{1u}} &=&S_{E_{1u}}\langle \cos ^{2}(\theta
_{E_{1u}})\rangle v  \label{Eqsmp2} \\
\widetilde{S}_{A_{2u}} &=&S_{A_{2u}}\langle \cos ^{2}(\theta
_{A_{2u}})\rangle v.  \label{Eqsmp3}
\end{eqnarray}%
Here, $v=0.7$ is the relative density of the unoriented pressed pellet
compared with that of the single crystal.\cite{Experiments,NoteDensity} A
benchmark is now needed to related the apparent and intrinsic oscillator
parameters.

Fortunately, the polarized infrared reflectance of a 2H-MoS$_{2}$ single
crystal has been studied by Wieting \textit{et al}.\cite{Wieting1971} Fits
to the reflectance yield $\epsilon (\infty )_{E_{1u}}=15.2$, $\epsilon
(\infty )_{A_{2u}}=6.2$, $S_{E_{1u}}=0.20$, and $S_{A_{2u}}=0.03$ (Table \ref%
{Tab1}). These intrinsic parameters are appropriate benchmarks for the our
pellet samples, because they are made of $\mu m$ sized powders, for which
the surface effect can be ignored as a good approximation.

In Eqs. (\ref{Eqsmp1} - \ref{Eqsmp3}), the only unknowns are $\langle \cos
^{2}(\theta _{E_{1u}})\rangle $ and $\langle \cos ^{2}(\theta
_{A_{2u}})\rangle $. Fig. \ref{fig1}(b) shows a schematic view of the
pressed pellet of 2H- powder. Since the polarizations of the two modes are
orthogonal, one has $\theta _{E_{1u}}+\theta _{A_{2u}}=\frac{\pi }{2}$.
Hence, $\langle \cos ^{2}(\theta _{E_{1u}})\rangle +\langle \cos ^{2}(\theta
_{A_{2u}})\rangle =1$, meaning that there is only one unknown, say, $\langle
\cos ^{2}(\theta _{E_{1u}})\rangle $. Using Eqs. (\ref{Eqsmp2}-\ref{Eqsmp3}%
), we find two independent values as 0.81 and 0.83 in good agreement with
each other. To further check the self-consistency, we calculated $\widetilde{%
\epsilon }(\infty )$ using Eq. (\ref{Eqsmp1}) using an average value of $%
\langle \cos ^{2}(\theta _{E_{1u}})\rangle $ = 0.82, yielding $\widetilde{%
\epsilon }(\infty )=9.8$, in excellent agreement with that obtained by
direct fitting techniques ($\widetilde{\epsilon }(\infty )=10.3$). 
In addition to confirming the validity of our approach, this
self-consistency also shows that the pressed powder sample of 2H-MoS$_{2}$
will have the same Born effective charge as the single crystal,\cite%
{Wieting1971,Sun2009} which, of course, it must. The dielectric
response $\varepsilon_1$ was calculated using  the intrinsic
parameters of Ref. \onlinecite{Sun2009} and is plotted in the inset
of Fig. \ref{fig1}(b).   The dispersive response is typical of an
anisotropic insulator with two vibrational modes.

The availability of $\sim $30 - 70 nm average diameter nested MoS$_{2}$
nanoparticles provides an opportunity to investigate the impact of finite
length scale effects on chemical bonding.\cite{Sun2009} Figure \ref{fig1}%
(c) displays a close-up view of the measured far infrared reflectance along
with an oscillator fit. The apparent parameters obtained from this fit can
be scaled toward a set of intrinsic oscillator parameters using the
orientation and density corrections outlined above.\cite{Sun2009} From an
analysis of these intrinsic oscillator strengths and high frequency
dielectric constants along with mode frequencies, we can extract Born and
local effective charges for the nanoparticles.\cite{Sun2009} In the
intralayer direction, we find that the Born effective charge of the
nanoparticles is 0.69 $e$ in the intralayer direction, significantly lower
than that of the layered bulk (1.11 $e$). Here, $e$ is the charge of an
electron. We attribute this difference to structural strain (and resulting
change in intralayer polarizability) in the nanoparticles.\cite{Sun2009}
The Born effective charge of the nanoparticles remains unchanged in
the interlayer direction (0.52 $e$). The dielectric constant was
again  calculated using intrinsic parameters\cite{Sun2009} and is
plotted in the inset of Fig. \ref{fig1}(c). Clearly, the dispersive
response of the nanoparticles is different from that of the bulk in
the intralayer direction. They are the same in the interlayer
direction.

Extension of Born and local (or ionic) charge concepts to nanomaterials is
an important advance because most are not (and may never be) available in an
oriented form.\cite%
{Long2007,Tenne2004,Poudel2008,Hochbaum2008,Boukai2008,Cao2007}
Indeed, emerging mechanical and tribological applications of
nanoscale MoS$_2$ require bulk quantities of powder with careful
size-shape control but no orientational control. At the same time,
the relationship
between engineering properties such as solid state lubrication\cite%
{Rapoport1997,Seifert2000,Remskar2007} and the microscopic aspects of charge
and bonding is an open and interesting question that deserves further study.

\section{Conclusion}

We present an application of the Lorentz model in which fits to vibrational
spectra or a Kramers Kronig analysis of the reflectance are employed along
with several useful formalisms to quantify microscopic charge and
polarizability in unoriented (powdered) materials. This paper provides a
systematic development of the operative equations and a discussion of the
conditions under which such techniques can be employed. We demonstrate the
workability of our approach by analyzing the vibrational response of a
layered transition metal dichalcogenide, and we include an evaluation of
Born and local (or ionic charge) of its nanoscale analog to illustrate the
modern utility. The extension to assess size effects advances the field of
nanoscience and, at the same time, retains many attractive features of
optical spectroscopy and the traditional Lorentz model.

\section*{Acknowledgments}

This work was supported by the Joint Directed Research and
Development Program at the University of Tennessee and Oak Ridge
National Laboratory. We thank R. Tenne for the high resolution TEM
image of the nanoparticles  and many useful discussions.

\appendix* 

\section{Other frameworks}

For applications, it can be convenient to use  other forms of Eq.
(\ref{Eqdi}) that are written in terms of parameters that are more
related to the experimental observations. Before stepping into that,
we take a careful look at Eq. (\ref{Eqdi}) and note that there are
only three independent parameters: $\alpha $, $q$, and $\omega
_{0}$. Other three-variable sets offer equivalent expressions. Some
of these are detailed below.

\subsection{$\protect\varepsilon (0)$, $\protect\varepsilon (\infty )$ and $%
\protect\omega _{TO}^{2}$ framework}

This is a very useful framework because $\varepsilon (0)$,
$\varepsilon (\infty )$, and $\omega _{TO}^{2}$ can all be
straightforwardly extracted  from the optical spectra.

\subsubsection{Collinear Oscillators}

Equation (\ref{EqExpTh}) is already very close to employing this new
set of parameters. Let's define:

\end{subequations}
\begin{eqnarray}
\varepsilon (0) &=&(1+\frac{\alpha /V}{1-\eta \alpha /V})+\frac{q^{2}}{%
\epsilon _{0}mV(1-\eta \alpha /V)^{2}}\frac{1}{\omega _{TO}^{2}}  \notag \\
&=&\varepsilon (\infty )+\frac{q^{2}}{\epsilon _{0}mV(1-\eta \alpha /V)^{2}}%
\frac{1}{\omega _{TO}^{2}}
\end{eqnarray}%
Then,
\begin{equation*}
A=\frac{q^{2}}{mV(1-\eta \alpha /V)^{2}}=\omega _{TO}^{2}[\varepsilon
(0)-\varepsilon (\infty )]
\end{equation*}%
with parameters $\varepsilon (0)$, $\varepsilon (\infty )$, and
$\omega _{TO}^{2}$, one can rewrite Eq. (\ref{Eqdi}):

\begin{equation}
\varepsilon =\varepsilon (\infty )+\omega _{TO}^{2}\frac{\varepsilon
(0)-\varepsilon (\infty )}{\omega _{TO}^{2}-\omega ^{2}-i\gamma \omega }
\label{Eqdi2}
\end{equation}


\begin{itemize}
\item $\omega _{0}^{2}$
\end{itemize}

It is a bit tedious, but not so difficult to show that:

\begin{equation*}
\omega _{0}^{2}=\omega _{TO}^{2}\frac{\varepsilon (\infty )+1/\eta -1}{%
\varepsilon (0)+1/\eta -1}
\end{equation*}

\begin{itemize}
\item Ionic effective charge
\end{itemize}

\begin{eqnarray}
q^{2} &=&\epsilon _{0}mV(1-\eta \alpha /V)^{2}A  \notag \\
&=&\epsilon _{0}mV(1-\eta \alpha /V)^{2}\omega _{TO}^{2}[\varepsilon
(0)-\varepsilon (\infty )]  \notag \\
&=&\epsilon _{0}mV\frac{\omega _{TO}^{2}[\varepsilon (0)-\varepsilon (\infty
)]}{\eta ^{2}[\varepsilon (\infty )+1/\eta -1]^{2}}
\end{eqnarray}

\begin{itemize}
\item Born effective charge
\end{itemize}

\begin{equation}
q_{B}^{2}=\epsilon _{0}mV\omega _{TO}^{2}[\varepsilon (0)-\varepsilon
(\infty )]
\end{equation}

\begin{itemize}
\item $\omega _{LO}^{2}$
\end{itemize}

By definition,  $\omega _{LO}$, $\varepsilon =0$  for a longitudinal
mode. Using
Eq. (\ref{Eqdi2}) (and ignoring $i\gamma \omega $), one has %
\begin{equation}
0=\varepsilon (\infty )+\omega _{TO}^{2}\frac{\varepsilon (0)-\varepsilon
(\infty )}{\omega _{TO}^{2}-\omega _{LO}^{2}}
\end{equation}%
which gives the Lyddane-Sachs-Teller relation,\cite{Ashcroft1976}%
\begin{equation}
\frac{\omega _{LO}^{2}}{\omega _{TO}^{2}}=\frac{\varepsilon (0)}{\varepsilon
(\infty )}  \label{EqLST}
\end{equation}

One can write out $\omega _{LO}^{2}$ in terms of microscopic parameters $%
\alpha $, $q$ and $\omega _{0}$ using the Lyddane-Sachs-Teller relation as follows:%
\begin{eqnarray}
\omega _{LO}^{2} &=&\omega _{TO}^{2}\frac{\varepsilon (0)}{\varepsilon
(\infty )}  \notag \\
&=&\omega _{TO}^{2}\frac{\varepsilon (\infty )+A\frac{1}{\omega _{TO}^{2}}}{%
\varepsilon (\infty )}  \notag \\
&=&\omega _{TO}^{2}+\frac{A}{\varepsilon (\infty )}  \notag \\
&=&\omega _{0}^{2}-\frac{q^{2}\eta }{\epsilon _{0}mV(1-\eta \alpha /V)}
\notag \\
&&+\frac{q^{2}}{\epsilon _{0}mV(1-\eta \alpha /V)^{2}}\frac{1}{1+\frac{%
\alpha /V}{1-\eta \alpha /V}}  \notag \\
&=&\omega _{0}^{2}+\frac{q^{2}}{\epsilon _{0}mV}\frac{1-\eta }{1+(1-\eta
)\alpha /V}
\end{eqnarray}%
Note that $\omega _{LO}^{2}$ is large than $\omega _{0}^{2}$, while $\omega
_{TO}^{2}$ is smaller than $\omega _{0}^{2}$.

\subsubsection{Multiple Collinear Oscillators}

\begin{equation}
\varepsilon =1+\underset{j}{\sum }[\varepsilon _{j}(\infty )-1]+\underset{j}{%
\sum }\omega _{TO,j}^{2}\frac{\varepsilon _{j}(0)-\varepsilon _{j}(\infty )}{%
\omega _{TO,j}^{2}-\omega ^{2}-i\gamma _{j}\omega }
\end{equation}

\subsubsection{Tilted Oscillators}

\begin{eqnarray}
\varepsilon &=&1+[\varepsilon(\infty )-1)]\langle\cos ^{2}(\theta )\rangle
\notag \\
&+&\omega _{TO}^{2}\frac{\varepsilon (0)-\varepsilon (\infty )}{%
\omega_{TO}^{2}-\omega ^{2}-i\gamma \omega } \langle\cos ^{2}(\theta )\rangle
\end{eqnarray}

\subsubsection{Multiple Tilted Oscillators}

\begin{eqnarray*}
\varepsilon (\{\theta _{j}\}) &=&1+\underset{j}{\sum }[\varepsilon
_{j}(\infty )-1]\cos ^{2}(\theta _{j}) \\
&&+\underset{j}{\sum }\omega _{TO,j}^{2}\frac{\varepsilon
_{j}(0)-\varepsilon _{j}(\infty )}{\omega _{TO,j}^{2}-\omega ^{2}-i\gamma
_{j}\omega }\cos ^{2}(\theta _{j})
\end{eqnarray*}

\begin{align}
\widetilde{\varepsilon }& =\langle \varepsilon (\{\theta _{j}\})\rangle
\notag \\
& =1+\underset{j}{\sum }[\varepsilon _{j}(\infty )-1]\langle \cos
^{2}(\theta _{j})\rangle   \notag \\
& +\underset{j}{\sum }\omega _{TO,j}^{2}\frac{\varepsilon
_{j}(0)-\varepsilon _{j}(\infty )}{\omega _{TO,j}^{2}-\omega ^{2}-i\gamma
_{j}\omega }\langle \cos ^{2}(\theta _{j})\rangle
\end{align}

\subsection{$\protect\varepsilon (\infty )$, $\protect\omega _{TO}^{2}$ and $%
\protect\omega _{LO}^{2}$ framework}

\subsubsection{Collinear oscillators}

Another choice is to use $\varepsilon (\infty )$, $\omega _{TO}^{2}$, and $%
\omega _{LO}^{2}$. Using the Lyddane-Sachs-Teller relation, Eq. (\ref{EqLST}), one can eliminate $%
\varepsilon (0)$, which yields%
\begin{equation}
\varepsilon =\varepsilon (\infty )+\varepsilon (\infty )\frac{\omega
_{LO}^{2}-\omega _{TO}^{2}}{\omega _{TO}^{2}-\omega ^{2}-i\gamma \omega }
\label{Eqdi3}
\end{equation}

\begin{itemize}
\item Ionic effective charge
\end{itemize}

\begin{eqnarray}
q^{2} &=&\epsilon _{0}mV(1-\eta \alpha /V)^{2}A  \notag \\
&=&\epsilon _{0}mV(1-\eta \alpha /V)^{2}\varepsilon (\infty )(\omega
_{LO}^{2}-\omega _{TO}^{2})  \notag \\
&=&\epsilon _{0}mV\frac{\varepsilon (\infty )(\omega _{LO}^{2}-\omega
_{TO}^{2})}{\eta ^{2}[\varepsilon (\infty )+1/\eta -1]^{2}}
\end{eqnarray}

\begin{itemize}
\item Born effective charge
\end{itemize}

\begin{equation}
q_{B}^{2}=\epsilon _{0}mV\varepsilon (\infty )(\omega _{LO}^{2}-\omega
_{TO}^{2})
\end{equation}

\subsubsection{Multiple Collinear Oscillators}

\begin{equation}
\varepsilon =1+\underset{j}{\sum }[\varepsilon _{j}(\infty )-1]+\underset{j}{%
\sum }\varepsilon _{j}(\infty )\frac{\omega _{LO,j}^{2}-\omega _{TO,j}^{2}}{%
\omega _{TO,j}^{2}-\omega ^{2}-i\gamma _{j}\omega }
\end{equation}

\subsubsection{Tilted Oscillators}

\begin{eqnarray}
\varepsilon &=&1+[\varepsilon(\infty )-1]\langle\cos ^{2}(\theta )\rangle
\notag \\
&+&\varepsilon (\infty )\frac{\omega _{LO}^{2}-\omega _{TO,j}^{2}}{%
\omega_{TO}^{2}-\omega ^{2}-i\gamma \omega } \langle\cos ^{2}(\theta )\rangle
\end{eqnarray}

\subsubsection{Multiple Tilted Oscillators}

\begin{align}
\varepsilon (\{\theta _{j}\})& =1+\underset{j}{\sum }[\varepsilon
_{j}(\infty )-1]\cos ^{2}(\theta _{j})  \notag \\
& +\underset{j}{\sum }\varepsilon _{j}(\infty )\frac{\omega
_{LO,j}^{2}-\omega _{TO,j}^{2}}{\omega _{TO,j}^{2}-\omega ^{2}-i\gamma
_{j}\omega }\cos ^{2}(\theta _{j})
\end{align}

\begin{align}
\widetilde{\varepsilon }& =\langle \varepsilon (\{\theta _{j}\})\rangle
\notag \\
& =1+\underset{j}{\sum }[\varepsilon _{j}(\infty )-1]\langle \cos
^{2}(\theta _{j})\rangle   \notag \\
& +\underset{j}{\sum }\varepsilon _{j}(\infty )\frac{(\omega
_{LO,j}^{2}-\omega _{TO,j}^{2})}{\omega _{TO,j}^{2}-\omega ^{2}-i\gamma
_{j}\omega }\langle \cos ^{2}(\theta _{j})\rangle   \label{final}
\end{align}


\begin{references}

\bibitem{Born1918}  M. Born, Phys. Z. {\bf19}, 539 (1918).


\bibitem{Szigeti1950}   B. Szigeti, Proc. R. Soc. London A {\bf204}, 51 (1950).

\bibitem{Ashcroft1976} N. W. Ashcroft, and N. D. Mermin, \textit{Solid State Physics} (Thomson Learning, New York, 1976).

\bibitem{Kittel1966}    C. Kittel, \textit{Introduction to Solid State Physics} (Wiley, New York, 1966).

\bibitem{Born1954}  M. Born and K. Huang, \textit{Dynamic Theory of Crystal Lattices} (Oxford University Press, London, 1954).

\bibitem{Lee2003}   K. W. Lee and W. E. Pickett, Phys. Rev. B {\bf68}, 085308 (2003).

\bibitem{Fontana1984}   M. D. Fontana, G. Metrat, J. L. Servoin, and F. Gervais, J. Phys. C {\bf17}, 483 (1984).

\bibitem{Zhong1994} W. Zhong, R.D. King-Smith, and D. Vanderbilt, Phys. Rev. Lett. {\bf 72}, 3618 (1994).

\bibitem{Long2007} J.W. Long and D.R. Rolison, Acc. Chem. Res. {\bf40}, 854 (2007).

\bibitem{Tenne2004} R. Tenne and C. N. R. Rao, Phil. Trans. R. Soc. A {\bf362}, 2099 (2004).

\bibitem{Poudel2008}    B. Poudel, Q. Hao, Y. Ma, Y. C. Lan, A. Minnich, B. Yu, X. Yan, D. Z. Wang, A. Muto, D. Vashaee, X. Y. Chen, J. M. Liu, M. S. Dresselhaus, G. Chen, and Z. Ren, Science {\bf320}, 634 (2008).

\bibitem{Hochbaum2008} A. I. Hochbaum, R. K. Chen, R. D. Delgado, W. J. Liang, E. C. Garnett, M. Najarian, A. Majumdar, and P. D. Yang, Nature {\bf451}, 163 (2008).

\bibitem{Boukai2008}    A. I. Boukai, Y. Bunimovich, J. Tahir-Kheli, J. K. Yu, W. A. Goddard, and J. R. Heath, Nature {\bf451}, 168 (2008).

\bibitem{Cao2007} J. Cao, J.L. Musfeldt, S. Mazumdar, N.A. Chernova, and M.S. Whittingham, Nano. Lett. {\bf 7}, 2351 (2007).

\bibitem{Carr1985} G.L. Carr, S. Perkowitz, and D.B. Tanner, Infrared and Millimeter Waves {\bf 13}, 177 (1985).


\bibitem{Sun2009}   Q.-C. Sun, X. S. Xu, L. I. Vergara, R. Rosentsveig, and J. L. Musfeldt, Phys. Rev. B {\bf79}, 205405 (2009).

\bibitem{Wieting1971}   T. J. Wieting and J. L. Verble, Phys. Rev. B {\bf3}, 4286 (1971).

\bibitem{Uchida1978} S. I. Uchida and S. Tanaka, J. Phys. Soc. Jpn {\bf45}, 153 (1978).

\bibitem{Wooten1972}  F. Wooten, \textit{Optical Properties of Solids} (New York, Academic Press, 1972).

\bibitem{Huang1988} K. Huang and R. Q. Han, \textit{Solid State Physics} (in Chinese) (Higher Education Press, China, 1988).

\bibitem{Gonze1997} X. Gonze and C. Lee, Phys. Rev. B {\bf55}, 10355 (1997).


\bibitem{Schonfeld1983} B. Schonfeld, J. J. Huang and S. C. Moss,  Acta Crystallogr. Sect. B \textbf{39}, 404 (1983).

\bibitem{Verble1970}    J. L. Verble and T. J. Wieting, Phys. Rev. Lett. {\bf25}, 362 (1970).

\bibitem{Experiments} IF-MoS$_2$ was prepared as described
previously.\cite{Sun2009}  2H-MoS$_{2}$ was purchased directly from
Alfa Aesar (99\%). Pressed pellet samples were prepared using low
pressure (see Fig. \ref{fig1}(a)). The theoretical density of
2H-MoS$_{2}$ single crystal is 4.996 g/cm$^{3}$, and the actual
densities of 2H- and IF- pellets are $\sim $3.5 g/cm$^{3}$. Pellet
densities are therefore $\sim $70\% of the single crystal density, a
difference that we correct for in our analysis. We checked that
there is no significant infrared transmittance of the samples
(thickness between 1.0 and 1.3 mm), 
indicating large attenuation due to strong resonance.
Near normal infrared reflectance was measured using a Bruker 113V
Fourier transform infrared spectrometer. A helium-cooled bolometer
detector was employed in the far-infrared for added sensitivity.
Both 0.5 and 2 cm$^{-1}$ resolution were used in the
infrared.\cite{Sun2009}

\bibitem{NoteDensity} For far infrared spectroscopy, the wavelength of the
light is usually much larger than the particle size. Therefore, the
measured oscillator strength is proportional to the concentration of
the oscillators, which is inversely proportional to the mass density
of the powder sample.



\bibitem{Rapoport1997} L. Rapoport, Y. Bilik, Y. Feldman, M. Homyonfer, S. R. Cohen, and R. Tenne, Nature \textbf{387}, 791 (1997).

\bibitem{Seifert2000} G. Seifert, H. Terrones, M. Terrones, G. Jungnickel, and T. Frauenheim, Phys. Rev. Lett. \textbf{85}, 146 (2000).

\bibitem{Remskar2007} M. Remskar, A. Mrzel, M. Virsek, and A. Jesih, Adv. Mater. \textbf{19}, 4276 (2007).




\end{references}
\end{document}